\documentclass[12 pt,twoside,aps,prd,amsmath,amssymb,
tightenlines,showpacs,showkeys,eqsecnum]{revtex4}
\usepackage{mathptmx}
\usepackage{bm}
\usepackage{graphicx}
\usepackage[colorlinks]{hyperref}

\newcommand {\ve}{\varepsilon}

\newcommand {\cG}{\cal G}
\newcommand {\cD}{\cal D}
\newcommand {\G}{\Gamma}

\newcommand {\bg}{\bar \gamma}

\newcommand {\bp}{\bar \psi}

\begin{document}
\title{Spinor fields in spherically symmetric space-time}
\author{Bijan Saha}
\affiliation{Laboratory of Information Technologies\\
Joint Institute for Nuclear Research, Dubna\\
141980 Dubna, Moscow region, Russia\\ and\\
Institute of Physical Research and Technologies\\
People's Friendship University of Russia\\
Moscow, Russia} \email{bijan@jinr.ru}
\homepage{http://spinor.bijansaha.ru}

\hskip 1 cm

\begin{abstract}
Within the scope of a spherically symmetric space-time we study the
role of a nonlinear spinor field in the formation of different
configurations with spherical symmetries. The presence of the
non-diagonal components of energy-momentum tensor of the spinor
field leads to some severe restrictions on the spinor field itself.
Since spinor field is the source of the gravitational one,  the
metric functions also changes in accordance with it. The system as a
whole possesses solutions only in case of some additional conditions
on metric functions.
\end{abstract}



\maketitle

\section{Introduction}

In the recent past spinor description of matter and dark energy was
used to draw the picture of the evolution of the Universe within the
scope of Bianchi type anisotropic cosmological models
\cite{1997GRG,2001PRD,2006PRD,2018EChAYa}. It was found that the
approach in question gives rise to a variety of solutions depending
on the choice of spinor field nonlinearity. Thanks to its
sensitivity to gravitational field spinor field brings some
unexpected nuances in the behavior of both the spinor and the
gravitational fields. Taking this in mind in this paper we consider
the nonlinear spinor field within the framework of spherically
symmetric gravitational field. Since a variety of astrophysical
systems such as stars, black holes are described by spherically
symmetric configurations, the use of spinor field in this area might
be very promising.

\section{Basic Equation}

Let us consider a system of nonlinear spinor and spherically
symmetric gravitational fields. The corresponding action we choose
in the form

\begin{eqnarray}
{\cal S}(g; \psi, \bp) = \int\, \left(L_{\rm g} + L_{\rm sp}\right)
\sqrt{-g} d\Omega \label{action}
\end{eqnarray}

Here $L_{\rm g}$ corresponds to the gravitational field

\begin{eqnarray}
L_{\rm g} = \frac{R}{2\kappa}, \label{lgrav}
\end{eqnarray}

where  $R$ is the scalar curvature,  $\kappa = 8 \pi G$ with $G$
being Newton's gravitational constant and $L_{\rm sp}$ is the spinor
field Lagrangian which we take in the form

\begin{equation}
L_{\rm sp} = \frac{\imath}{2} \left[\bp \gamma^{\mu} \nabla_{\mu}
\psi- \nabla_{\mu} \bar \psi \gamma^{\mu} \psi \right] - m \bp \psi
- F, \label{lspin}
\end{equation}

with the nonlinear term $F = F(K)$ and $K$ taking one of the
following expressions: $\{I,\,J,\,I+J,\,I-J\}$. Here $I = \bp \psi$
and $J = \imath \bp \bg^5 \psi$. Here $m$ is the spinor mass.

The spinor field equations corresponding to the spinor field
Lagrangian \eqref{lspin} are

\begin{subequations}
\label{speq}
\begin{eqnarray}
\imath\gamma^\mu \nabla_\mu \psi - m \psi - {\cD} \psi -
 \imath {\cG} \gamma^5 \psi &=&0, \label{speq1} \\
\imath \nabla_\mu \bp \gamma^\mu +  m \bp + {\cD}\bp + \imath {\cG}
\bp \gamma^5 &=& 0, \label{speq2}
\end{eqnarray}
\end{subequations}
where we denote ${\cD} = 2 S F_K K_I$ and ${\cG} = 2 P F_K K_J$,
with $F_K = dF/dK$, $K_I = dK/dI$ and $K_J = dK/dJ.$ In view of
\eqref{speq} it can be shown that
\begin{equation}
L_{\rm sp} = 2 K F_K - F. \label{LspinF}
\end{equation}

In the above expressions $\nabla_\mu \psi = \partial_\mu \psi -
\G_\mu \psi$ and $\nabla_\mu \bp = \partial_\mu \bp + \bp \G_\mu$
with $\G_\mu$ being the spinor affine connection.

The spherically-symmetric metric we choose in the form
\begin{equation}
ds^2 = e^{2 \mu} dt^2 - e^{2 \alpha} dr^2 - e^{2 \beta}
(d\vartheta^2 + \sin^2{\vartheta} d \varphi^2), \label{ss}
\end{equation}
where the metric functions $\mu, \alpha, \beta$ depend on the
spatial variable $r$ only.

The spinor affine connection matrices are defined as
\begin{equation}
\Gamma_\mu (x)= \frac{1}{4}g_{\rho\sigma}(x)\biggl(\partial_\mu
e_{\delta}^{b}e_{b}^{\rho} -
\Gamma_{\mu\delta}^{\rho}\biggr)\gamma^\sigma\gamma^\delta,
\label{gm}
\end{equation}

where the tetrad $e_b^\rho$ correspond to the metric \eqref{ss}  we
choose as follows:

\begin{equation}
e_0^{(0)} = e^\mu, \quad e_1^{(1)} = e^\alpha, \quad e_2^{(2)} =
e^\beta, \quad e_3^{(3)} = e^\beta \sin \theta. \label{tetradss}
\end{equation}

The flat $\gamma$ matrices we choose in the from
\begin{eqnarray}
\bg^0 &=& \left(\begin{array}{cc}I & 0 \\ 0 & -I\end{array}\right),
\quad
\bg^1 = \left(\begin{array}{cc}0 & \sigma^1\\
-\sigma^1 & 0 \end{array}\right),\nonumber \\
\bg^2 &=& \left(\begin{array}{cc}0 & \sigma^2\\
-\sigma^2 & 0 \end{array}\right), \quad
\bg^3 = \left(\begin{array}{cc}0 & \sigma^3\\
-\sigma^3 & 0 \end{array}\right). \nonumber
\end{eqnarray}

where
\begin{eqnarray}
I &=& \left(\begin{array}{cc}1 & 0 \\ 0 & 1\end{array}\right), \quad
\sigma^1 = \left(\begin{array}{cc}\cos{\vartheta} & \sin{\vartheta} e^{-\imath \varphi}\\
\sin{\vartheta} e^{\imath \varphi} &
-\cos{\vartheta}\end{array}\right), \nonumber\\
\sigma^2 &=&
\left(\begin{array}{cc} - \sin{\vartheta} & \cos{\vartheta} e^{-\imath \varphi}\\
\cos{\vartheta} e^{\imath
\varphi}&\sin{\vartheta}\end{array}\right), \quad \quad \sigma^3 =
\left(\begin{array}{cc}0 &\imath e^{-\imath \varphi} \\-\imath
e^{\imath \varphi} & 0\end{array}\right).
\end{eqnarray}

Defining $\gamma^5$ as follows:
\begin{eqnarray}
\gamma^5&=&-\frac{i}{4} E_{\mu\nu\sigma\rho}\gamma^\mu\gamma^\nu
\gamma^\sigma\gamma^\rho, \quad E_{\mu\nu\sigma\rho}= \sqrt{-g}
\ve_{\mu\nu\sigma\rho}, \quad \ve_{0123}=1,\nonumber \\
\gamma^5&=&-i\sqrt{-g} \gamma^0 \gamma^1 \gamma^2 \gamma^3
\,=\,-i\bg^0\bg^1\bg^2\bg^3 = \bg^5, \nonumber
\end{eqnarray}
we obtain
\begin{eqnarray}
\bg^5&=&\left(\begin{array}{cc}0&-I\\-I&0
\end{array}\right).\nonumber
\end{eqnarray}

Taking into account that the functions $\alpha$, $\beta$ and $\mu$
depend on only $r$ ($x^1$) from \eqref{gm} we find

\begin{subequations}
\label{bgn}
\begin{eqnarray}
\G_0 &=& - \frac{1}{2} \mu^\prime\, e^{(\mu -\alpha)}\,\bg^0 \bg^1,
\label{bg0n}\\
\G_0 &=& 0, \label{bg1n}\\
\G_2 &=&\frac{1}{2} \beta^\prime\, e^{(\beta -\alpha)}\,\bg^2 \bg^1,
\label{bg2n} \\
\G_3 &=& \frac{1}{2} \beta^\prime\, \sin \theta e^{(\beta
-\alpha)}\,\bg^3 \bg^1 + \frac{1}{2} \cos \theta \bg^3 \bg^2.
\label{bg3n}
\end{eqnarray}
\end{subequations}

Let us now find the energy-momentum tensor of the spinor field which
is given by

\begin{eqnarray}
T_{\mu}^{\,\,\,\rho}&=&\frac{\imath}{4} g^{\rho\nu} \left(\bp
\gamma_\mu \nabla_\nu \psi + \bp \gamma_\nu \nabla_\mu \psi -
\nabla_\mu \bar \psi \gamma_\nu \psi - \nabla_\nu \bp \gamma_\mu
\psi \right) \,- \delta_{\mu}^{\rho} L_{\rm sp} \nonumber\\
&=& \frac{\imath}{4} g^{\rho\nu} \left(\bp \gamma_\mu
\partial_\nu \psi + \bp \gamma_\nu \partial_\mu \psi -
\partial_\mu \bar \psi \gamma_\nu \psi - \partial_\nu \bp \gamma_\mu
\psi \right)\nonumber\\
& - &\frac{\imath}{4} g^{\rho\nu} \bp \left(\gamma_\mu \G_\nu +
\G_\nu \gamma_\mu + \gamma_\nu \G_\mu + \G_\mu \gamma_\nu\right)\psi
 \,- \delta_{\mu}^{\rho} L_{\rm sp}.
\label{temspApp}
\end{eqnarray}

On account of spinor field equations one finds the following
non-trivial and linearly independent terms of the energy-momentum
tensor
\begin{subequations}
\label{spemt}
\begin{eqnarray}
T^0_0 &=& F(K) - 2 K F_K, \label{T00}\\
T^1_1 &=& m S + F(K),  \label{T11}\\
T^2_2 &=& F(K) - 2 K F_K,  \label{T22}\\
T^3_3 &=& F(K) - 2 K F_K, \label{T3}\\
T^0_1 &=& \frac{1}{4}\frac{\cos \theta}{\sin \theta} e^{(\alpha - \mu -\beta)} A^3, \label{T01}\\
T^0_2 &=& -\frac{1}{4} \left(\mu^\prime -
\beta^\prime\right)\,e^{(\beta -\alpha - \mu)} A^3, \label{T02}\\
T^0_3 &=&  \frac{1}{4} \left(\mu^\prime - \beta^\prime\right) \,
e^{( \beta -\alpha - \mu)} \sin \theta A^2 +\frac{1}{4}  e^{-\mu}
\cos \theta  A^1. \label{T03}
\end{eqnarray}
\end{subequations}

We consider the case when the spinor field depends on $r$ only. Then
in view of \eqref{bgn} we have

\begin{subequations}
\label{speqn}
\begin{eqnarray}
\imath e^{-\alpha} \bg^1 \psi^\prime
+\frac{\imath}{2}\left(\mu^\prime + 2 \beta^\prime \right)
e^{-\alpha}\bg^1 \psi + \frac{\imath}{2} \frac{\cos \theta}{\sin
\theta} e^{-\beta} \bg^2 \psi  - m \psi - {\cD}
\psi -  \imath {\cG} \gamma^5 \psi &=&0, \label{speq1n} \\
\imath e^{-\alpha} \bp^\prime \bg^1 +
\frac{\imath}{2}\left(\mu^\prime + 2 \beta^\prime \right)
e^{-\alpha} \bp \bg^1 + \frac{\imath}{2} \frac{\cos \theta}{\sin
\theta} e^{-\beta}\bp \bg^2  +  m \bp + {\cD}\bp + \imath {\cG} \bp
\gamma^5 &=& 0. \label{speq2n}
\end{eqnarray}
\end{subequations}

For the invariants of spinor field from \eqref{speqn} we find

\begin{subequations}
\label{spinv}
\begin{eqnarray}
S^\prime + \left( \mu^\prime + 2 \beta^\prime\right) S - 2 e^\alpha
{\cG} A^1 &=& 0, \label{Sinv}\\
P^\prime + \left( \mu^\prime + 2 \beta^\prime\right) P + 2 e^\alpha
\left( m + {\cD}\right) A^1 &=& 0, \label{Pinv}\\
{A^1}^\prime + \left( \mu^\prime + 2 \beta^\prime\right) A^1 +
\frac{\cos \theta}{\sin \theta}  e^{\alpha - \beta} A^2 + 2 e^\alpha
\left( m + {\cD}\right) P +  2 e^\alpha
{\cG} S &=& 0, \label{A1inv}\\
{A^2}^\prime + \left( \mu^\prime + 2 \beta^\prime\right) A^2 -
\frac{\cos \theta}{\sin \theta}  e^{\alpha - \beta} A^1 &=& 0,
\label{A2inv}
\end{eqnarray}
\end{subequations}
where $A^\mu = \bp \gamma^5 \gamma^\mu \psi$ is the pseudovector.
The foregoing system gives
\begin{eqnarray}
\left(S S^\prime - P P^\prime + A^1 {A^1}^\prime + A^2
{A^2}^\prime\right) + \left( \mu^\prime + 2 \beta^\prime\right)
\left(S^2 - P^2 + {A^1}^2 + {A^2}^2\right) = 0, \label{Invrel}
\end{eqnarray}
with the relation

\begin{eqnarray}
\left(S^2 - P^2 + {A^1}^2 + {A^2}^2\right) = C_0 e^{-2\left( \mu + 2
\beta\right)}, \label{Invrel0}
\end{eqnarray}

On account of \eqref{spemt} we find the following system of Einstein
equations

\begin{subequations}
\label{EET}
\begin{eqnarray}
\left( 2 \mu^\prime \beta^\prime +
\beta^{\prime 2}\right) - e^{2 (\alpha - \beta)} &=& - \kappa e^{2 \alpha} \left(m S + F(K)\right), \label{EE11}\\
\left(\mu^{\prime 2} + \mu^\prime \beta^\prime -\mu^\prime
\alpha^\prime + \beta^{\prime 2} -\beta^\prime \alpha^\prime +
\mu^{\prime\prime} +
\beta^{\prime\prime}\right) &=& -\kappa e^{2 \alpha} \left(F(K) - 2 K F_K\right), \label{EE22}\\
\left(3 \beta^{\prime 2} - 2 \beta^\prime \alpha^\prime +
2\beta^{\prime\prime}\right)- e^{2 (\alpha - \beta)} &=& - \kappa
e^{2 \alpha} \left(F(K)
- 2 K F_K\right), \label{EE00}\\
0 &=& \frac{\cos \theta}{\sin \theta} e^{(\alpha - \mu -\beta)} A^3, \label{EE01}\\
0 &=& \left(\mu^\prime -
\beta^\prime\right)\,e^{(\beta -\alpha - \mu)} A^3, \label{EE02}\\
0 &=& \left[ \left(\mu^\prime - \beta^\prime\right) \, e^{(\beta
-\alpha)}  A^2 +  \frac{ \cos \theta}{\sin \theta}  A^1\right].
\label{EE03}
\end{eqnarray}
\end{subequations}

From \eqref{EE01} we obtain $A^3 = 0$ at least everywhere expect
$\theta = \pi/2.$ Hence \eqref{EE02} fulfills identically.

Inserting $  \frac{ \cos \theta}{\sin \theta} e^{(\alpha - \beta)}
A^1 =  - \left( \mu^\prime - \beta^\prime\right) A^2$ from
\eqref{EE03} into \eqref{A2inv} for $A^2$ we find

\begin{equation}
{A^2}^\prime + \left(2\mu^\prime +  \beta^\prime\right) A^2  = 0,
\label{A2inv1}
\end{equation}
with the solution
\begin{equation}
A^2 = z_2 e^{-\left(2\mu +  \beta\right)}, \label{A2}
\end{equation}
with $z_2$ being some integration constant.

On account of \eqref{spemt} from Bianchi identity
$\G^{\nu}_{\mu;\nu} = 0$ i.e.,
\begin{equation}
T^{\nu}_{\mu;\nu} = T^{\nu}_{\mu,\nu} +  \G^{\nu}_{\alpha\nu}
T^{\alpha}_{\mu} - \G^{\alpha}_{\mu\nu} T^{\nu}_{\alpha} = 0
\label{BIden}
\end{equation}
we find
\begin{eqnarray}
\left(m S + F\right)^\prime +  \left( \mu^\prime + 2 \beta^\prime
\right)  \left(m S + 2 K F_K\right) = 0. \label{TOVgen1}
 \end{eqnarray}

Let us now consider two different cases. In case of $K = I = S^2$,
on account of $F^\prime = 2 S F_K S^\prime$ from \eqref{TOVgen1} we
find

\begin{eqnarray}
\left(m  + 2 S F_K\right) S^\prime + \left(\mu^\prime +  2
\beta^\prime\right)\left(m S + 2 S^2 F_K\right) = 0, \label{TOV1}
\end{eqnarray}
which leads to
\begin{equation}
S^\prime + \left(\mu^\prime  + 2 \beta^\prime\right) S = 0,
\label{TOgenV2}
\end{equation}
with
\begin{equation}
S = c_1 e^{-(\mu + 2 \beta)}. \label{Sgen}
\end{equation}
In case of $K$ one of $\{J,\,I+J,\,I-J\}$ we consider the massless
spinor field, as it was done in cosmology \cite{1997GRG,2001PRD}.
Then on account of $F^\prime = F_K K^\prime$ we rewrite
\eqref{TOVgen1} as
\begin{eqnarray}
 F_K K^\prime + 2 \left( \mu^\prime + 2 \beta^\prime\right) K F_K = 0, \label{TOVgen1n}
\end{eqnarray}
which leads to
\begin{equation}
K^\prime + 2 \left(\mu^\prime  + 2 \beta^\prime\right) K = 0,
\label{TOVgen2n}
\end{equation}
with
\begin{equation}
K = c_1^2 e^{-2(\mu + 2 \beta)}. \label{Kgen}
\end{equation}
Thus we conclude that the relations \eqref{Kgen} holds for a
massless spinor field if $K$ takes one of $\{I,\,J,\,I+J,\,I-J\}$,
whereas it is true for a non-trivial spinor mass, only if $K = I$.

Now we can deal with the zero component of $A^\mu$. From Fierz
identity we know

\begin{eqnarray}
S^2 +  P^2 = - A_\mu A^\mu = -\left({A^0}^2 + {A^1}^2 + {A^2}^2 +
{A^3}^2\right), \label{Invrel0a}
\end{eqnarray}

Subtraction of \eqref{Invrel0a} from \eqref{Invrel0} in view of $A^3
= 0$ leads to

\begin{eqnarray}
{A^0}^2  = - C_0 e^{-2\left( \mu + 2 \beta\right)} - 2 P^2,
\label{Invrel0b}
\end{eqnarray}

whereas their addition yields

\begin{eqnarray}
{A^0}^2  =  C_0 e^{-2\left( \mu + 2 \beta\right)} - 2 \left(S^2 +
{A^1}^2 + {A^2}^2 \right). \label{Invrel0c}
\end{eqnarray}

Hence all the components of $A^\mu$ can be expressed in terms of
metric functions.

Thus the non-diagonal Einstein equations together with the equations
for invariants of spinor field give us valuable information about
the spinor field. On the other hand Bianchi identity relates the
invariants with metric functions. Now we have only three diagonal
Einstein equations left.

Before dealing with diagonal Einstein equations let us go back to
spinor field equations. As far as the spinor field equation
\eqref{speq1} is concerned, denoting $\phi = \psi e^{(\mu + 2
\beta)/2}$ it can be rewritten in the following matrix form

\begin{eqnarray}
\phi^\prime = B \phi, \label{speq1n}
\end{eqnarray}

where

\begin{eqnarray}
B = \left(\begin{array}{cc}Y_3 \sigma^1 + \imath Y_1 \sigma^3 & Y_2 \sigma^1\\
- Y_2\sigma^1 & -Y_3 \sigma^1 + \imath Y_1 \sigma^3
\end{array}\right),
\quad \phi = {\rm col}\left(\phi_1,\,\phi_2,\,\phi_3,\,\phi_4\right), \nonumber \\
, \label{epeq1n}
\end{eqnarray}
and $Y_1 = \left(\cos{\vartheta}/2\sin{\vartheta}\right)
\exp{[\alpha - \beta]}$,\,\, $Y_2 = \imath \left[m + {\cD}\right]
\exp{\alpha}$, and $Y_3 = {\cG} \exp{\alpha}$.

As one sees, $\det{B} = - Y_1^2 + Y_2^2 -Y_3^2 \ne 0$. The solution
to the equation \eqref{speq1n} can be written in general form.

Let us now go back to Einstein field equations. In order to solve
these equations we need to consider some special cases. In what
follows we consider a few of them.

{\bf Case I}  Let us assume that

\begin{eqnarray}
\alpha = \mu + 2 \beta. \label{addcon}
\end{eqnarray}

This type of assumption was consider in a number of papers
\cite{bronnikov1979,SahaIJTP1997}  and known as harmonic condition.

Denoting $ \beta^{\prime 2} + \beta^\prime \mu^\prime = U $  we
rewrite the diagonal equations of Einstein system, i.e.
\eqref{EE11}, \eqref{EE22} and \eqref{EE00} as follows:

\begin{subequations}
\label{EETdh}
\begin{eqnarray}
e^{-2 \alpha} U - e^{-2 \beta} &=& - \kappa \left(m S + F\right) \label{EE11h}\\
e^{-2 \alpha}\left(\mu^{\prime \prime} + \beta^{\prime \prime}
- U \right) &=& \kappa \left(2 K F_K - F\right) \label{EE22h}\\
e^{-2 \alpha}\left(2\beta^{\prime\prime} - U\right)- e^{-2 \beta}
&=& \kappa \left(  2 K F_K - F\right) \label{EE00h}
\end{eqnarray}
\end{subequations}

Subtraction of \eqref{EE00h} from \eqref{EE22h} gives

\begin{eqnarray}
\mu^{\prime \prime} - \beta^{\prime \prime} + e^{2(\mu + \beta)} =
0, \label{22h-00h}
\end{eqnarray}

Subtraction of \eqref{EE00h} from \eqref{EE11h} gives

\begin{eqnarray}
\beta^{\prime \prime} - \beta^{\prime 2} - \beta^{\prime} \mu^\prime
= \frac{\kappa}{2}  e^{2(\mu + 2\beta)} \left(m S + 2 K F_K \right)
\label{11h-00h}
\end{eqnarray}

As far as F is concerned, we may choose it in the form $F = \lambda
K^n$, where $\lambda$ is the self-coupling constant. In case of $ K
= I = S^2$ we consider a massive spinor field, otherwise massless
one.

$ K = I = S^2$ we find the following system of equations

\begin{subequations}
\begin{eqnarray}
\mu^{\prime \prime} - \beta^{\prime \prime} &=& - e^{2(\mu + \beta)}
\label{22h-00h1}\\
\beta^{\prime \prime} - \beta^{\prime 2} - \beta^{\prime} \mu^\prime
&=& \frac{\kappa}{2} \bigl(c_1 m e^{(\mu + 2\beta)}  + 2 \lambda n
c_1^{2n} e^{2(1-n)(\mu + 2\beta)}  \bigr)  \label{11h-00h1}
\end{eqnarray}
\end{subequations}

\begin{figure}[ht]
\centering
\includegraphics[width=50mm]{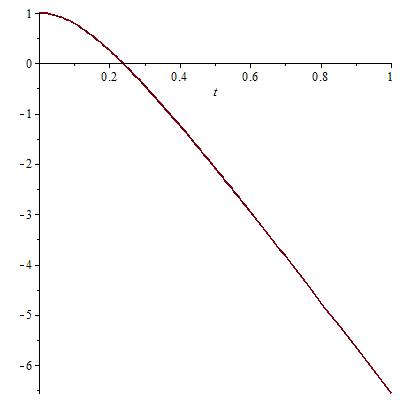}\\
\caption{Behavior of $\mu(r)$ for $\lambda=1,\, n=1,\, m=1,\,
\kappa=1,\, c_1=1$ with the initial conditions $\mu(0)=1,\,
\beta(0)=1,\, \mu^\prime(0)=0, \beta^\prime(0)=0$}
\label{mucase1n1}.
\end{figure}

\begin{figure}[ht]
\centering
\includegraphics[width=50mm]{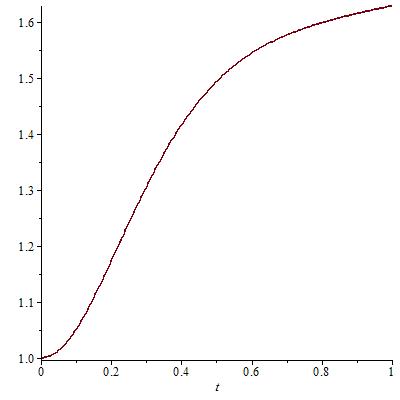}\\
\caption{Behavior of $\beta(r)$ for $\lambda=1,\, n=1,\, m=1,\,
\kappa=1,\, c_1=1$ with the initial conditions $\mu(0)=1,\,
\beta(0)=1,\, \mu^\prime(0)=0, \beta^\prime(0)=0$}
\label{betacase1n1}.
\end{figure}

\begin{figure}[ht]
\centering
\includegraphics[width=50mm]{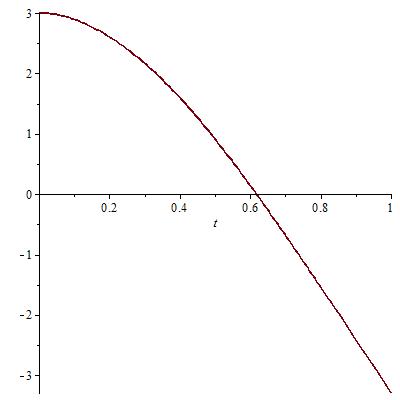}\\
\caption{Behavior of $\alpha(r)$ for $\lambda=1,\, n=1,\, m=1,\,
\kappa=1,\, c_1=1$ with the initial conditions $\mu(0)=1,\,
\beta(0)=1,\, \mu^\prime(0)=0, \beta^\prime(0)=0$}
\label{alphacase1n1}.
\end{figure}

The foregoing system we solved numerically. In doing so we have give
some concrete value of problem parameters as well as initial
conditions. For simplicity we have chosen $\lambda=1,\, n=1,\,
m=1,\, \kappa=1,\, c_1=1$. As initial conditions we have set
$\mu(0)=1,\, \beta(0)=1,\, \mu^\prime(0)=0, \beta^\prime(0)=0$. Here
our main aim was to find some solutions which we can use in our
following detailed studies. In Figs. \ref{mucase1n1},
\ref{betacase1n1} and \ref{alphacase1n1} we have plotted the metric
functions $\mu(r), \beta(r)$ and $\alpha(r)$, respectively.

{\bf Case II} As a second case we consider the widely used model
setting $ \beta = \ln r$. In this case the diagonal components of
Einstein system takes the form

\begin{subequations}
\label{EETdg}
\begin{eqnarray}
e^{-2 \alpha}\left( 2 \frac{\mu^\prime}{r} +
\frac{1}{r^2}\right) - \frac{1}{r^2} &=& - \kappa \left(m S + F\right) \label{EE11dg}\\
e^{-2 \alpha}\left(\mu^{\prime 2} + \frac{\mu^\prime}{r} -
\mu^\prime \alpha^\prime  - \frac{\alpha^\prime}{r} +
\mu^{\prime\prime}
\right) &=& \kappa \left(2 K F_K - F\right) \label{EE22dg}\\
e^{-2 \alpha}\left(\frac{1}{r^2} - 2 \frac{\alpha^\prime}{r}\right)-
\frac{1}{r^2} &=&  \kappa \left(2 K F_K - F\right) \label{EE00dg}
\end{eqnarray}
\end{subequations}

Subtracting \eqref{EE00dg} from  \eqref{EE22dg} one finds

\begin{eqnarray}
 e^{-2\alpha} \left(\mu^\prime + \alpha^\prime\right) = - \frac{\kappa
r}{2} \left(m S + 2 K F_K\right). \label{11-00}
\end{eqnarray}

Inserting \eqref{11-00} into \eqref{EE22dg} we find

\begin{eqnarray}
e^{-2 \alpha}\left(2 \mu^{\prime 2} + 2\frac{\mu^\prime}{r} +
\mu^{\prime\prime} \right) &= & \kappa \left(2 K F_K - F\right) -
\frac{\kappa r}{2} \left(\mu^\prime + \frac{1}{r}\right)\left(m S +
2 K F_K\right). \label{EE22dgn}
\end{eqnarray}

It should be noted that in this case $S = (c_1/r^2) e^{-\mu}$. As
far as nonlinear term is concerned, as in previous case we consider
the massive spinor field with $F = \lambda I^n = \lambda S^{2n}$.

Inserting $F = \lambda I^n = \lambda S^{2n}$ into the equations
finally we find the following system

\begin{subequations}
\begin{eqnarray}
e^{-2\alpha} \left(\mu^\prime + \alpha^\prime\right) &=& -
\frac{\kappa r}{2} \left(\frac{m c_1}{r^2} e^{-\mu} + \frac{2
\lambda
n c_1^{2n}}{r^{4n}} e^{-2 n \mu}\right) \label{11-001}\\
e^{-2 \alpha} \left(2 \mu^{\prime 2} + 2\frac{\mu^\prime}{r} +
\mu^{\prime\prime} \right) &=&  \kappa \left(\frac{(2n - 1)\lambda
c_1^{2n}}{r^{2n}} e^{-2\mu}\right) \nonumber\\ &-& \frac{\kappa
r}{2}\left(\mu^\prime + \frac{1}{r}\right) \left(\frac{m c_1}{r^2}
e^{-\mu} + \frac{2 \lambda n c_1^{2n}}{r^{4n}} e^{-2 n
\mu}\right)\label{EE22dgn1}
\end{eqnarray}
\end{subequations}

\begin{figure}[ht]
\centering
\includegraphics[width=50mm]{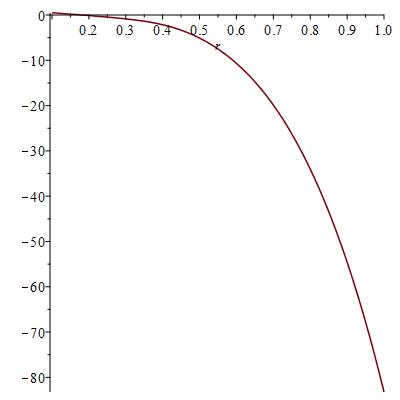}\\
\caption{Behavior of $\mu(r)$ for $\lambda=1,\, n=1,\, m=1,\,
\kappa=1,\, c_1=1$ with the initial conditions $\mu(0.1)=0.5,\,
\alpha(0.1)=1,\, \mu^\prime(0.1)=0.2$} \label{mucase2n1}.
\end{figure}

\begin{figure}[ht]
\centering
\includegraphics[width=50mm]{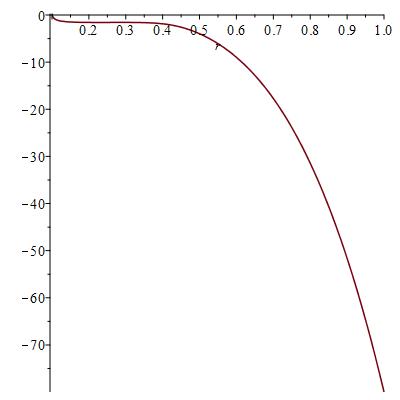}\\
\caption{Behavior of $\alpha(r)$ for $\lambda=1,\, n=1,\, m=1,\,
\kappa=1,\, c_1=1$ with the initial conditions $\mu(0.1)=0.5,\,
\alpha(0.1)=1,\, \mu^\prime(0.1)=0.2$} \label{alphacase2n1}.
\end{figure}

This system can also be solved numerically to find $\mu$ and
$\alpha$. Since in this case the point $r = 0$ leads to singularity,
we have to set the initial value at any point except this one. Like
in the previous we consider the same problem parameters with the
following initial conditions: $\alpha (0.1) = 0.3,\, \mu (0.1) =
0.5,$ and $\mu^\prime (0.1) = 0.2$. The behavior of $\mu (r)$ and
$\alpha (r)$ are given in the Figs. \ref{mucase2n1} and
\ref{alphacase2n1}, respectively. It should be noted that this case
was studied in \cite{Krechet}

\section{Conclusion}

Within the scope of spherically symmetric gravitational the the role
of nonlinear spinor field in the formation of different
configuration is studied. Earlier it was found that the spinor field
play a very important rope in the evolution of the Universe. This
study can be taken as an attempt to exploit the spinor field in
astrophysics. We hope to use these experience and results for the
well studied astrophysical objects such as compact stars, black
holes, wormholes etc. We plan to discus some aspects of modern
astrophysics exploiting the spinor description of matter in our
coming papers.

 \section*{Acknowledgments}

This work is supported in part by a joint Romanian-LIT, JINR, Dubna
Research Project, theme no. 05-6-1119-2014/2018.

\end{document}